\documentclass{pasj00}
\draft

\begin{document}
\SetRunningHead{T. Yamagami and Y. Fujita}{Star Formation in the Tail}

\title{Star Formation in the Cometary Tails Associated with Cluster
Galaxies}

\author{Takahiro \textsc{Yamagami} %
  \thanks{Present Address: Graduate School of 
Engineering Science, Osaka University, 
1-1 Machikaneyama-cho, Toyonaka, Osaka 560-0043,
email: t-yamagami@mit.eng.osaka-u.ac.jp}
and Yutaka \textsc{Fujita}
}
\affil{Department of Earth and Space Science, 
Graduate School of Science, Osaka University, \\
1-1 Machikaneyama-cho, Toyonaka, Osaka 560-0043}

%

\KeyWords{galaxies: active --- galaxies: clusters: general --- ISM: clouds --- stars: formation} 

\maketitle

\begin{abstract}
 We investigate the star-formation in cometary tails of galaxies in
 clusters. In particular, we focus on the evolution of molecular clouds
 in the tails that generate the stars. Assuming that the gas tails had
 been derived from the galaxies through ram-pressure stripping, we found
 that the gas must have been stripped mostly not in the form of
 molecular clouds, but in the form of H\emissiontype{I} gas or molecular
 gas that is not in clouds. Moreover, the molecular clouds are condensed
 in the tails even away from the host galaxies. We also found that
 magnetic fields may be required to suppress Kelvin-Helmholtz (KH)
 instability on the surface of molecular clouds, because otherwise KH
 instability may destroy the molecular clouds before stars are formed in
 them.
\end{abstract}

\section{Introduction}
\label{sec:intro}

Recent observations of nearby clusters of galaxies have shown that some
galaxies in the clusters have cometary tails (e.g.
\cite{gav01,ken04,yos04,mac05,wan04,oo05,sun05,chu07,cor07}). In the
optical band, for example, \citet{yag07} found an long and narrow ($\sim
60\times 2$~kpc) tail associated with the galaxy D~100 in the Coma
cluster. The relative velocity between the tail and the host galaxy is
$\sim 100$--$300\rm\: km\: s^{-1}$. \citet{yos08} discovered an unusual
complex of the narrow blue filaments derived from the galaxy RB~199 in
the Coma cluster and bright knots in these filaments.  These knots are
found to be young stars. Their masses are $\sim 10^6$--$10^7\:M_\odot$
and their sizes are $\sim 200$~pc. Judging from their colors, they are
estimated to be formed $\sim 10^8$ yr ago. The distance between the host
galaxy and the knots is $\gtrsim 10\: $~kpc \citep{yos08}.  In the X-ray
band, \citet{fuj06b} found a comet-like structure around one of the
brightest galaxies in Abell~2670. \citet{sun10} also found the X-ray
tails associated with the spiral galaxy ESO~137-001 in Abell~3627. They
confirmed active star-formation in the tails even at positions $\gtrsim
10$~kpc away from the host galaxy. In the infrared band, \citet{siv10}
discovered a molecular hydrogen tail with an H$_2$ mass of approximately
$4\times 10^7\: M_\odot$ extending 20~kpc from ESO~137-001.

It is often believed that the tails are created through ram-pressure
stripping of interstellar medium of galaxies, which has been intensively
studied \citep{gun72,tak84,gae87,fuj99}. Recently, detailed numerical
simulations of the ram-pressure stripping and formation of the tails
have been performed
\citep{aba99,ste99,mor00,qui00,sch01,bek03,roe05,roe06,kap09,ton10}.
However, although the resolution of the numerical simulations has
increased, it is still difficult to resolve physical processes on a
scale of molecular clouds. For example, in current simulations on
galactic scales, it would be difficult to resolve Kelvin-Helmholtz (KH)
instability, which is expected to develop around a molecular cloud
moving along the galactic tail.

In this paper, we study the formation of the galactic tails and
star-formation in them. We consider disk galaxies with abundant gas as
the host galaxies. We especially focus on the molecular clouds in which
the stars were born. We estimate the growth time of KH instability
around the clouds and compare it with the time-scale of star-formation
in the clouds, which is affected by the ram-pressure on them. For these
purposes, we assume that the molecular clouds are gravitationally
bound and we use the virial theorem for the clouds.

\section{Origin of the Molecular Clouds in Tails}
\label{sec:origin}

In this section, we discuss the formation process of the tails through
ram-pressure stripping. Since we are interested in the star-formation in
the tails, we argue whether the molecular clouds that produce stars in
the tails were formed in the tails or in the host galaxy.

\subsection{Ram-Pressure Stripping}
\label{sec:ram}

Although the number density of intracluster medium (ICM) is low
($\rho_{\rm ICM}/m_{\rm H} \sim 10^{-3} \rm\:cm^{-3}$; $m_{\rm H}$ is
the mass of a hydrogen atom), ram-pressure on a galaxy is relatively
large, because the galaxy moves very fast (typically $\gtrsim 1000\: \rm
km\:s^{-1}$) in a cluster. The ram-pressure is given by
\begin{equation}
\label{eq:ramP}
P_{\rm ram}= \rho_{\rm ICM}v_{\rm gal}^2=1.7\times 10^{-11} 
{\rm dyn\: cm^{-2}}
\left(\frac{\rho_{\rm ICM}}{10^{-3}\: m_{\rm H}\:\rm cm^{-3}}\right)
\left(\frac{v_{\rm gal}}{10^3\rm\: km\: s^{-1}}\right)^2 \:,
\end{equation}
where $v_{\rm gal}$ is the velocity of the galaxy relative to the
ICM. On the other hand, the gravity from the galaxy per unit area is
given by
\begin{equation}
\label{eq:fg}
 f_{\rm g} = 2 \pi G \Sigma_{\rm s} \Sigma_{\rm gas}
= v_{\rm rot}^{2}R_{\rm gal}^{-1} \Sigma_{\rm gas} \:,
\end{equation}
where $G$ is the gravitational constant, $\Sigma_{\rm s}$ is the
gravitational surface mass density of the galaxy, $v_{\rm rot}$ is the
rotation velocity, and $R_{\rm gal}$ is the radius of the disk
\citep{gun72,fuj99}. The column density of the gas we consider is given
by $\Sigma_{\rm gas}$.
Most numerical simulations have implicitly considered ram-pressure
stripping of H\emissiontype{I} gas with a density of $\sim 1\rm\:
cm^{-3}$. As their results have been shown, the H\emissiontype{I} gas
can easily be stripped. In fact, if we replace $\Sigma_{\rm gas}$ by the
typical value of the gas column density of the disk, $\Sigma_{\rm HI}$
(e.g. \cite{bin87}), it is written as
\begin{eqnarray}
f_{\rm g} &=& 2.1\times 10^{-11} 
{\rm dyn\: cm^{-2}}\left(\frac{v_{\rm rot}}{220\rm\: km\:
					     s^{-1}}\right)^{2}
\nonumber\\
& &\times\left(\frac{R_{\rm gal}}{10\rm\: kpc}\right)^{-1}
 \left(\frac{\Sigma_{\rm HI}}
{8\times 10^{20}\: m_{\rm H}\rm\: cm^{-2}}\right) 
\end{eqnarray}
and is comparable to $P_{\rm ram}$ in equation~(\ref{eq:ramP}). Thus, if
$v_{\rm gal}$ is larger than $\sim 1000\rm\: km\: s^{-1}$, the gas is
stripped.

Next, we consider the ram-pressure stripping of a molecular cloud in a
galaxy, which is expect to be more difficult than that of
H\emissiontype{I} gas, because of its high density \citep{bos06}. In
this case, $\Sigma_{\rm gas}$ in equation~(\ref{eq:fg}) should be
replaced by the column density of the molecular cloud.  The virial
theorem for a molecular cloud gives the relation between the mass
$M_{\rm c}$, the radius $R_{\rm c}$, and the external pressure $P_{\rm
c}$ of the cloud:
\begin{equation}
\label{eq:virial} 
\frac{M_{\rm c}}{R_{\rm c}^2}= a \:
\left(\frac{P_{\rm c}}{10^4\: k_{\rm B} \: \rm cm^{-3}\: K} 
\right)^{1/2} \;,
\end{equation}
where $a=190 \pm 90\: M_\odot {\rm pc^{-2}}$ and $k_{\rm B}$ is the
Boltzmann constant \citep{elm89}. If we define the column density of the
cloud as $\Sigma_{\rm c} = M_{\rm c}/(\pi R_{\rm c}^2)$ and assume that
the cloud is affected by the ram-pressure ($P_{\rm c}=P_{\rm ram}$), the
column density is given by
\begin{equation}
\label{eq:sigc}
 \Sigma_{\rm c} = 2.6\times 10^{22}\: m_{\rm H}\:{\rm cm^{-2}}
\left(\frac{P_{\rm ram}}{1.7\times 10^{-11} 
{\rm dyn\: cm^{-2}}}\right)^{1/2}\:.
\end{equation}
Thus, from equation~(\ref{eq:fg}) we obtain
\begin{eqnarray}
\label{eq:fgc}
f_{\rm g} &\sim & 6.9\times 10^{-10} 
{\rm dyn\: cm^{-2}}\left(\frac{v_{\rm rot}}{220\rm\: km\:
					     s^{-1}}\right)^{2}
\nonumber\\
& &\times\left(\frac{R_{\rm gal}}{10\rm\: kpc}\right)^{-1}
 \left(\frac{\Sigma_{\rm c}}
{2.6\times 10^{22}\: m_{\rm H}\rm\: cm^{-2}}\right)\:.
\end{eqnarray}
Equations~(\ref{eq:ramP}) and (\ref{eq:fgc}) show that the ram-pressure
cannot overcome the gravity from the galaxy ($P_{\rm ram}\ll f_{\rm
g}$). The larger ram-pressure means the larger column density of a
molecular cloud [equation~(\ref{eq:sigc})], which tend to prevent
ram-pressure stripping of the cloud [equation~(\ref{eq:fgc})].

It is to be noted, however, there seem to be galaxies in which
molecular clouds have been stripped; NGC~4522 is an example
\citep{vol08}. For this galaxy, \citet{vol06} indicated that the galaxy
might have undergone extremely strong ram-pressure stripping. They
speculated that the relative velocity between the galaxy and the
surrounding ICM was larger because of the internal motion of the
ICM. For example, if we take $\rho_{\rm ICM}=4\times 10^{-3}\: m_{\rm
H}\rm\: cm^{-3}$ and $v_{\rm gal}=3000\rm\: km\: s^{-1}$, $P_{\rm ram}$
is comparable to $f_{\rm g}$ [equations~(\ref{eq:ramP}) and
(\ref{eq:fgc})]. If this was the case, the clouds could have been
stripped. Moreover, smaller gravity of a galaxy ($\propto v_{\rm
rot}^2/R_{\rm gal}$) makes ram-pressure stripping easier
[equation~(\ref{eq:fgc})]. Although ram-pressure stripping of molecular
clouds is difficult to happen, the observations may suggest that the
galaxy has been affected by exceptionally strong ram-pressure.

We also note that equation~(\ref{eq:virial}) is not valid when the
environment of a cloud rapidly changes. According to the virial theorem
for a cloud,
\begin{equation}
\label{eq:virial2}
\frac{c}{R_{\rm c}^{1/2}} \sim 0.4 \pm 0.1\: 
\Bigg(\frac{P_{\rm ram}}{10^4\: k_{\rm B}\: \rm cm^{-3}\: K} 
\Bigg) ^{1/4} \:\rm km\: s^{-1}\: pc^{-1/2}\;
\end{equation}
\citep{elm89}. For $R_{\rm c}=100$~pc and $P_{\rm ram}=1.7\times
10^{-11} {\rm dyn\: cm^{-2}}$, we obtain $c\sim 7.5\rm\: km\:
s^{-1}$. Thus, the typical time-scale of the evolution of the cloud is
$R_{\rm c}/c\sim 1\times 10^7$~yr. This may be compared with the
time-scale of ram-pressure stripping ($<10^8$~yr; e.g. \cite{qui00}). In
this subsection, however, we use equation~(\ref{eq:virial}) just to
check the condition of ram-pressure stripping. For that purpose,
equation~(\ref{eq:virial}) is appropriate enough, because we do not
consider the evolution of the cloud during the stripping. Moreover,
extremely large ram-pressure is required to strip the cloud as indicated
above. Thus, when the cloud is stripped, interstellar medium with lower
density has already been stripped from the galaxy, and the cloud
interacts directly with the ICM. Therefore, external pressure on the
cloud does not change much during the stripping.

In the following, we also consider molecular clouds in a tail away
from the host galaxy. For these clouds, the change of the environment is
very slow. For example, the crossing time of the host galaxy in a
cluster is $\sim 1$~Mpc/$1000\rm\: km\: s^{-1}\sim 1\times 10^9$~yr.
Therefore, the change of the environment does not affect the evolution
of the cloud and we can use equation~(\ref{eq:virial}).

\subsection{Migration of the Gas}
\label{sec:mig}

As mentioned in section~\ref{sec:intro}, stars are often found $\gtrsim
10$~kpc behind the host galaxies in clusters. Since each star is very
dense, it is obvious that it is not affected by ram-pressure. In other
words, ram-pressure cannot decelerate stars and draw them away from the
host galaxy. This means that the gas that makes up the stars at present
has been drawn from the galaxy not in the form of stars but in the form
of diffuse gas. The question is whether the gas was in the form of
H\emissiontype{I} gas or molecular clouds when it was migrating from the
host galaxy.

\begin{figure}
  \begin{center}
    \FigureFile(80mm,80mm){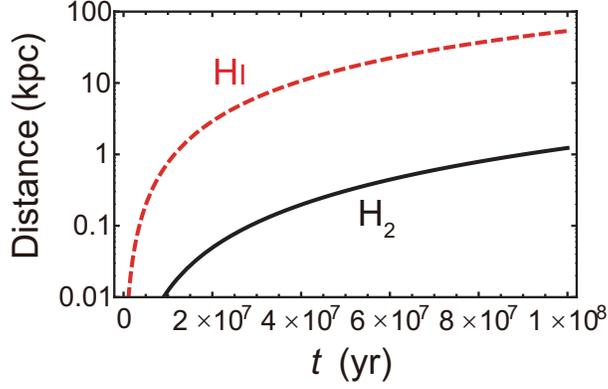}
  \end{center}
  \caption{Distance from the galaxy for H\emissiontype{I} gas (dashed)
 and a molecular cloud (solid). The ICM density is $\rho_{\rm
 ICM}/m_{\rm H} = 1\times 10^{-3}\rm\: cm^{-3}$ and the initial velocity
 of the gas is $v_{\rm c}(0)=v_{\rm gal}=2000\rm\: km\: s^{-1}$. The
 mass of the H\emissiontype{I} gas is $M_{\rm HI}=2\times 10^9\:
 M_\odot$. The result for the molecular cloud does not depend on its
 mass.}\label{fig:dist}
\end{figure}

First, we consider the migration of H\emissiontype{I} gas.  For
simplicity, we assume that the H\emissiontype{I} gas in the galactic
disk is stripped at one time, because the time-scale of the ram-pressure
stripping is very short ($<10^8$~yr; e.g. \cite{qui00}). The mass of the
H\emissiontype{I} gas blob is
\begin{eqnarray}
\label{HIstripping}
M_{\rm HI} &\sim& \Sigma_{\rm HI} \pi R_{\rm gal}^2 \nonumber\\
&=& 2\times 10^9\: M_\odot \left(\frac{\Sigma_{\rm HI}}
{8\times 10^{20}\: m_{\rm H}\:\rm cm^{-2}}\right)
\left(\frac{R_{\rm gal}}{10\rm\: kpc}\right)^{2} \:.
\end{eqnarray}
The equation of motion for the gas blob is
\begin{equation}
\label{eq:vHI}
 M_{\rm HI}\dot{v}_{\rm HI} \sim 
- \rho_{\rm ICM}v_{\rm HI}^2 \pi R_{\rm gal}^2 \:,
\end{equation}
where $v_{\rm HI}$ is the velocity of the blob relative to the ICM.  For
given $M_{\rm HI}$, $\rho_{\rm ICM}$, and $R_{\rm gal}$, one can solve
equation~(\ref{eq:vHI}) and obtain the distance between the blob and the
host galaxy, if it is assumed that the blob is detached from the galaxy
with the velocity of the galaxy, $v_{\rm gal}$, at $t=0$. It is to be
noted that a galaxy is not decelerated by ram-pressure on a time-scale
of $\gtrsim 10^9$~yr \citep{fuj06a}. For example, if we assume that
$M_{\rm HI}=2\times 10^9\: M_\odot$, $\rho_{\rm ICM}/m_{\rm H} = 1\times
10^{-3}\rm\: cm^{-3}$, $R_{\rm gal}=10\rm\: kpc$, and $v_{\rm
gal}=2000\rm\: km\: s^{-1}$, the distance between the blob and the
galaxy is $\sim 50$~kpc at $t\sim 1\times 10^8$~yr
(Figure~\ref{fig:dist}). The distance is comparable to the observed
length of tails (e.g. \cite{yos08,sun10}). Even at this time, the
velocity of the gas is still large ($v_{\rm HI}\sim 1000\rm\: km\:
s^{-1}$). In reality, the H\emissiontype{I} gas blob could decelerate
faster than the estimation above, because the blob should be shattered
through the interaction with the surrounding ICM, which increases the
surface area of the gas per unit mass.

Next, we consider the migration of a molecular cloud.  The equation of
motion for the cloud is
\begin{equation}
\label{eq:vc}
 M_{\rm c}\dot{v}_{\rm c} \sim 
- \rho_{\rm ICM}v_{\rm c}^2 \pi R_{\rm c}^2 \:,
\end{equation}
where $v_{\rm c}$ is the velocity of the cloud relative to the ICM. From
equation (\ref{eq:virial}), it can be easily shown that $\dot{v}_{\rm c}
\propto -\rho_{\rm ICM}^{1/2}v_{\rm c}$. Numerically, if we assume that
$\rho_{\rm ICM}/m_{\rm H} = 1\times 10^{-3}\rm\: cm^{-3}$ and $v_{\rm
c}(0)=v_{\rm gal}=2000\rm\: km\: s^{-1}$, the distance between the blob
and the galaxy is $\sim 1$~kpc even at $t\sim 1\times 10^8$~yr
(Figure~\ref{fig:dist}). Compared with the H\emissiontype{I} gas blob,
the distance is small, because the cloud is denser and the ram-pressure
per unit mass is smaller. Since the maximum age of a molecular cloud is
$\sim 10^8$~yr (section~\ref{sec:form}), this estimation indicates that
the cloud cannot travel the distance between the host galaxy and the
positions where star-formation has been observed ($\gtsim 10$~kpc;
e.g. \cite{yos08,sun10}).

In summary, the results we present in sections~\ref{sec:ram} and
\ref{sec:mig} show that in general the molecular clouds that bring forth
stars away from the host galaxy were formed neither inside nor close to
the host galaxy. In other words, dense part of the H\emissiontype{I} gas
that makes up the galactic tails condenses into molecular clouds even
away from the galaxy.

\section{Formation of Stars and Destruction of Clouds}
\label{sec:form}

\subsection{Kelvin-Helmholtz Instability}
\label{sec:KH}

The gas of the tail has a velocity comparable to that of the host galaxy
(e.g. \cite{yag07}). Thus, we expect that the molecular clouds that are
formed in the tail away from the galaxy also have large
velocities. Therefore, the interaction with the surrounding ICM could
lead to the development of KH instability around the clouds. Since
observations show that stars are born in the tails, KH instability must
be suppressed until the clouds turn into the stars. We do not consider
thermal conduction for simplicity (see section~\ref{sec:mag}).

If the self-gravity of a clouds is large enough, KH instability is
completely suppressed. The condition is
\begin{equation}
\label{eq:Mcr}
M_{\rm c}\gtrsim \frac{6^{1/2}\pi v_{\rm c}^3}
{G^{3/2}(\rho_{\rm c}/\rho_{\rm ICM})^2\rho_{\rm ICM}^{1/2}} \:,
\end{equation}
where $\rho_{\rm c}$ is the density of the molecular cloud and is given
by $\rho_{\rm c}=3 M_{\rm c}/(4\pi R_{\rm c}^3)$ \citep{mur93}.
Using equation~(\ref{eq:virial}) and assuming that $P_{\rm c}=\rho_{\rm
ICM}v_{\rm c}^2$, relation~(\ref{eq:Mcr}) can be written as
\begin{equation}
\label{eq:Mcr2}
 1 - 0.2\:\left(\frac{a}{190\: M_\odot {\rm pc^{-2}}}\right)^{-3}
\gtrsim 0 \:.
\end{equation}
Since $a=190 \pm 90\: M_\odot {\rm pc^{-2}}$ \citep{elm89}, the left
hand of relation~(\ref{eq:Mcr2}) can vary from -0.4 to 0.9. This means
that the cloud is at the boundary between stable and unstable states
and that the relation~(\ref{eq:Mcr}) is not a sufficient criterion of
stability in our case. Thus, we need to compare the growth time-scale of
KH instability ($\tau_{\rm KH}$) with the time-scale of star-formation in
the cloud ($\tau_{\rm form}$).

\subsection{Comparison of Time-Scales}

When KH instability is not completely suppressed, the growth time of the
instability is given by
\begin{equation}
 \label{tauKH}
\tau_{\rm KH}=\frac{(\rho_{\rm c}+\rho_{\rm ICM})R_{\rm c}}{(\rho_{\rm
c}\rho_{\rm ICM})^{1/2}v_{\rm c}}
\end{equation}
\citep{mur93}. Note that $R_{\rm c}$ is the function of $M_{\rm c}$ and
$P_{\rm c}$ [equation~(\ref{eq:virial})].

The star-formation in a molecular cloud under external pressure was
studied by \citet{elm97}. The equation for the rate of change of gas
mass in the cloud is
\begin{equation}
\label{eq:dMc/dt}
\frac{dM_{\rm c}}{dt}=-\frac{dM_{\rm s}}{dt}-\frac{AL}{c^2} \:,
\end{equation}
where $M_{\rm s}$ is the total mass of the embedded stars, $L$ is the
luminosity of the stars, $c$ is the velocity dispersion of the gas, and
$A$ is dimensionless constant. Equation~(\ref{eq:dMc/dt}) means that the
gas mass decreases because of the star-formation and the evaporation by
the stellar radiation. Assuming $L(t)$, and using
equation~(\ref{eq:virial}) and observational results to determine $d
M_{\rm s}/dt$ and $A$, one can solve equation~(\ref{eq:dMc/dt}), and
obtain the time when $M_{\rm c}(t)=0$ for given $M_{\rm c}(0)$ and
external pressure $P_{\rm c}$ \citep{elm97}. Since a significant
proportion of the molecular gas in the cloud converts to stars in this
time-scale, we define this time-scale as the time-scale of the
star-formation in the cloud ($\tau_{\rm form}$).

\begin{figure}
  \begin{center}
    \FigureFile(80mm,80mm){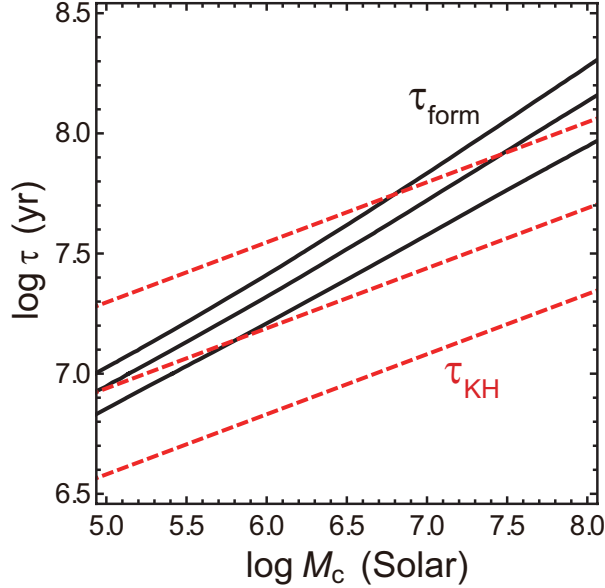}
  \end{center}
  \caption{Solid lines represent $\tau_{\rm form}$ and dashed lines
  represent $\tau_{\rm KH}$.  Each group has three lines corresponding
 to $v_{\rm c}=333, 1000, 3000 \:\rm km\: s^{-1}$ (from top to bottom).}
\label{fig:time}
\end{figure}

Figure~\ref{fig:time} shows $\tau_{\rm KH}$ and $\tau_{\rm form}$ for
molecular clouds with various masses $M_{\rm c}$ [$M_{\rm c}(0)$ for
$\tau_{\rm form}$] and velocities $v_{\rm c}$. We fixed $\rho_{\rm
ICM}/m_{\rm H}=1\times 10^{-3}\rm\: cm^{-3}$ and $a=190\: M_\odot {\rm
pc^{-2}}$ and gave the external pressure by $P_{\rm c}=\rho_{\rm
ICM}v_{\rm c}^2$. The formation time $\tau_{\rm form}$ in the figure
shows that a cloud disappears within $\lesssim 10^8$~yr even if there is
no KH instability. The comparison between $\tau_{\rm form}$ and
$\tau_{\rm KH}$ indicates that the cloud should be destroyed by KH
instability before forming stars if $v_{\rm c}\gtrsim 1000\rm\: km\:
s^{-1}$ ($\tau_{\rm KH}<\tau_{\rm form}$). We have confirmed that the
uncertainty of $a$ does not affect the conclusion. On the other hand,
although the velocities of stars formed in tails have not been
observationally studied well, the relative velocity of the tail gas to
the host galaxy is $\lesssim 200\rm\: km\: s^{-1}$ in the case of
ESO~137-001 \citep{sun10}. Thus, it is likely that stars are forming in
clouds with velocities of $v_{\rm c}\gtrsim 1000\rm\: km\: s^{-1}$.

One idea to overcome this contradiction is that clouds in tails take a
value of $a$ so that they are completely stable against KH instability
[equation~(\ref{eq:Mcr2})]. However, there is no solid ground for
that. Moreover, there may be some other uncertainties other than $a$,
because our estimation is simple. Therefore, we consider more persuasive
reason for the stability of a cloud in the next subsection.

\subsection{Magnetic Fields}
\label{sec:mag}

It is known that magnetic fields suppress KH instability. The interface
between the two magnetized fluids is stable if
\begin{equation}
B^2_{\rm ICM}+B^2_{\rm c} \geq 4\pi \frac{\rho_{\rm ICM} 
\rho_{\rm c}}{\rho_{\rm ICM}+\rho_{\rm c}}v_{\rm c}^2 \:,
\label{eq:tB}
\end{equation}
where $B_{\rm ICM}$ is the magnetic field of the ICM and $B_{\rm c}$ is
the one of a cloud \citep{lan60}. Even if the cloud is almost neutral,
minor ionization makes the relation~(\ref{eq:tB}) applicable
\citep{spi78}. Since $\rho_{\rm ICM}\ll \rho_{\rm c}$, the relation can
be written as
\begin{equation}
 \sqrt{B_{\rm ICM}^2+B_{\rm c}^2} \gtrsim 14\:\mu {\rm G}
\left(\frac{\rho_{\rm ICM}}{1\times 10^{-3}\: m_{\rm H}}\right)^{1/2}
\left(\frac{v_{\rm c}}{1000\:\rm km\: s^{-1}}\right) \:.
\end{equation}
The typical values of $B_{\rm ICM}$ and $B_{\rm c}$ are $\sim 3\: \mu\:
G$, and $\sim 30\: \mu\:G$, respectively
(e.g. \cite{sar86,spi78}). Thus, it is likely that magnetic fields
stabilize the surface of clouds in galactic tails even if their
self-gravity alone cannot. Moreover, if the fields are stretched along
the surface, thermal conduction will be suppressed \citep{vik01}.

We would like to note that similar phenomena may have been observed in
clusters. \citet{vik01} indicated that magnetic fields may suppress the
development of KH instability at contact discontinuities (cold fronts)
observed in ICM. They also pointed out that the magnetic fields may
suppress the thermal conduction across the discontinuity.

Our results show that molecular clouds with small velocities can survive
without magnetic fields. However, it is unlikely that only molecular
clouds with high velocities have magnetic fields. It may be natural to
assume that molecular clouds in the tails have similar magnetic fields
regardless of their velocities.

\section{Summary}

Recent observations have shown that some galaxies in clusters have long
tails ($>10$~kpc) and stars are forming in the tails.  We have studied
the formation of the tails and the star-formation in them, using the
virial theorem for gravitationally-bound molecular clouds. Our
scenario for the tail and star-formation can be summarized as follows.

\begin{itemize}

\item H\emissiontype{I} gas is stripped from a galaxy by
ram-pressure from the ICM. On the other hand, molecular clouds in the
galaxy are not directly stripped because
of the large gravity from the galaxy per unit surface area, unless
      the ram-pressure is extremely strong (section~\ref{sec:ram}).

\item While H\emissiontype{I} gas is easily decelerated by ram-pressure,
      molecular clouds are not, because they are denser. This means that
      the overall structure of the long tail is formed by the
       H\emissiontype{I}
      gas, and that molecular clouds that generate stars away from the
      host galaxy had condensed in the tail at positions away from the
       host galaxy. In other words, the clouds
      have not migrated from the host galaxy or from the neighbor of the
       host galaxy (section~\ref{sec:mig}).

\item For KH instability on their surfaces, the clouds are on the
       boundary between stable
      and unstable states, if only the self-gravity of the clouds is
       considered. However, if reasonable magnetic fields exist, they
       can
      suppress the development of KH instability
       (section~\ref{sec:form}).

\end{itemize}

\citet{fuj99b} (see also \cite{fuj98}) indicated that star-formation is
active in galaxies affected by ram-pressure until their interstellar
medium disappears. This will be applied to the galaxies with tails.  We
suppose that the galaxies with tails are the ones that have abundant gas
and fall into the central region of clusters for the first time. They
may be actively generating stars in the main bodies and the tails. This
is consistent with recent observations of galaxies in the Coma cluster
\citep{yag10}.

We speculate that the appearance of the tail would change as a galaxy
falls toward the cluster center. In the outer region of a cluster,
H\emissiontype{I} gas is stripped from an infalling galaxy and the tail
develops. As the galaxy reaches the central region of the cluster, some
of the H\emissiontype{I} gas cools and condenses into molecular
clouds. The rest of it will be mixed with the ICM. As the velocity of
the galaxy and the ICM density increases, the ram-pressure on the clouds
increases. Thus, the star-formation time-scale of the clouds is reduced
(Figure~\ref{fig:time}), and the star-formation becomes more efficient
until the clouds are consumed in the star-formation. H$\alpha$ emission
may be observed around this dense gas \citep{ton11}. Some of the
H\emissiontype{I} gas that was mixed with the ICM may be compressed by
the ambient ICM with high-pressure and may emit X-rays \citep{ton11}.

\bigskip

We thank the anonymous referee for useful comments. We are grateful to
F.~Takahara and T.~Tsuribe for useful discussion. This work was
supported by KAKENHI (20540269)

\end{document}